\documentstyle[12pt]{article}
\renewcommand\baselinestretch 2

\newenvironment{monenvir}[1]%
{\begin{center} \bf#1 \renewcommand\baselinestretch 1 \small}{\end{center}}

\begin{document}
\begin{center}
{\large\bf A SOLUTION TO THE HORIZON PROBLEM: \\
A DELAYED BIG-BANG SINGULARITY}
\end{center}

\bigskip

\begin{monenvir}
{Marie-No\"elle C\'EL\'ERIER$^{1}$, Jean SCHNEIDER$^{2}$} \\
D\'epartement d'Astrophysique Relativiste et de Cosmologie \\
Observatoire de Paris-Meudon \\
5 Place Jules Janssen 92195 Meudon C\'edex FRANCE\\
E-mail: $^1$ celerier@obspm.fr\\
      $^2$ schneider@obspm.fr \\
\bigskip
accepted for publication in Physics Letters A
\end{monenvir}

\bigskip
\noindent$\hrulefill$\\
\begin{abstract}
One of the main drawbacks of standard cosmology, known as the horizon 
problem, was until now thought to be only solvable in an inflationary 
scenario. A delayed Big-Bang in an inhomogeneous universe is shown to solve 
this problem while leaving unimpaired the main successful features
of the standard model.\\

\vskip1truecm
PACS: 98.80. Bp
\vskip1truecm

Keywords: horizon problem; delayed Big Bang; inhomogeneous universe;
Tolman-Bondi; cosmology

\end{abstract}

\medskip
\noindent$\hrulefill $

\bigskip
\section{Introduction}

Standard cosmology is known to rest on three observational pillars: the
expansion of the universe following Hubble law, the nearly isotropic black 
body  cosmic microwave background radiation (CMBR) and the abundances
of light elements produced during nucleosynthesis.

Besides these successfull predictions, it leaves ununderstood other peculiar
features of the observed universe.

In the present letter, a large class of initial singularity surfaces, the 
study of which has been initiated in a previous work \cite{SC98}, will 
be used to address one of the drawbacks of standard cosmology: the 
horizon problem.

The problem is the following: in hot Big Bang (BB) universes, the comoving 
region over which the CMBR is observed to be homogeneous to better than one 
part in $10^5$ is much larger than the comoving future light cone from the 
BB to the last scattering surface. The latter provides the maximal distance 
over which causal processes could have propagated since a given point on 
the BB surface. Hence, the observed quasi-isotropy of the CMBR remains 
unexplained.

Solving this problem was one of the main purposes of the inflationary paradigm
as it was first put forward by Brout, Englert and Gunzig \cite{BEG79} in 1979 
and independently by Guth \cite{G81} in 1981. But, as inflationary scenarii 
inflatened, some self-produced undesirable features came into the way: 
for instance, reheating is not actually well understood and important details 
are still under study \cite{K96}.

But despite these drawbacks, inflation has by now become a quasi-standard
paradigm as it was thought to be the only way to deal with the major
horizon problem.

Following a suggestion by Hu, Turner and Weinberg \cite{HTW94}, 
Liddle \cite{L95} has even proposed a proof that inflation is the only 
possible causal mechanism capable of generating density perturbations on 
scales well in excess of the Hubble radius, and hence the only way of solving 
the horizon problem.

As it was stressed by these authors, this problem involves the homogeneity 
and isotropy of the Freedmann-Robertson-Walker (FRW) model, proceeding from 
the Cosmological Principle upon which rests standard cosmology. Liddle's 
entire argument depends only on the properties of the FRW metric.

The so-called Cosmological Principle is in fact not an a priori principle, 
but at most a simplifying working hypothesis: the universe being as it is, all 
astrophysics can do is to build models compatible with observation, should 
they contradict the Cosmological Principle. We come back to this point in the 
discussion.

The purpose of this letter is to solve the horizon problem by means of a
delayed BB singularity in an inhomogeneous model of universe, thus discarding 
this Principle. For simplicity, we use a Tolman-Bondi model, since it allows 
a fully analytical exact reasoning. 

This model will be described in section 2. Calculations and arguments will be 
developed in section 3 and some examples given in section 4. Section 5 will 
be devoted to a brief discussion of the results and to the conclusion.

\section{An inhomogeneous delayed Big Bang model}

In an expanding universe, going backward along the parameter called the
cosmic time $t$ means going to growing energy densities and temperatures.

As one goes down the past, from our present matter dominated age defined
by the constant temperature hypersurface $T\sim 2.73^{\circ} K$, one 
reachs an epoch when the radiation energy density overcomes the matter one.
This radiation dominated area lasts until Planck time, $T_{Pl}\sim 10^{19}
GeV$, which marks the limit beyond which quantum gravitational
effects are expected to confuse our understanding of the laws of physics.

To deal with the horizon problem, one has to compute light cones. As will
be further shown, a large class of models can be found for which the
horizon problem is solved by means of light cones never leaving the matter
dominated area. We  will thus retain the Tolman-Bondi model for dust, an 
ideal non zero rest mass pressureless gas.

\subsection{The class of Tolman-Bondi models retained}

The Bondi line-element \cite{B47}, in comoving coordinates ($r,\theta,
\varphi$) and proper time $t$, is:
\begin{equation}
ds^2 = -c^2 dt^2 + S^2(r,t)dr^2 + R^2 (r,t)(d\theta^2 + \sin^2 \theta
d \varphi^2)  \label{eq:1}
\end{equation}

Solving Einstein's equation for this metric with the dust stress-energy
tensor gives:
\begin{eqnarray}
S^2(r,t) &=& {R^{'2}(r,t)\over 1+2E(r)/c^2} \label{eq:2}   \\
{1\over 2} \dot{R}^2(r,t) &-& {GM(r)\over R(r,t)}=E(r)  \label{eq:3}  \\
4\pi \rho (r,t) &=& {M'(r) \over R'(r,t) R^2(r,t)}   \label{eq:4}
\end{eqnarray}

where a dot denotes differentiation with respect to $t$ and a prime with
respect to $r$. $\rho (r,t)$ is the energy density of the matter.

$E(r)$ and $M(r)$ are arbitrary functions of $r$. $E(r)$ can be interpreted
as the total energy per unit mass and $M(r)$ as the mass within
the sphere of comoving radial coordinate $r$.

$M(r)$ remaining constant with time, it is used to define a radial
coordinate $r$: $M(r)\equiv M_0 r^3$, where $M_0$ is a constant.

Equation (\ref{eq:3}) can be solved and gives a parametric expression for
$R(r,t)$ for $E(r)\not= 0$ and an analytic one for $E(r)=0$.

As there are evidences that the observed universe does not present 
appreciable spatial curvature, it can be reliably approximated by a flat 
$E(r)=0$ Tolman-Bondi model.

With the above definition for the radial coordinate $r$, $R(r,t)$ possesses
thus an analytical expression, which we write:
\begin{equation}
R(r,t)=\left({9GM_0\over 2}\right)^{1/3} r[t-t_0(r)]^{2/3}  \label{eq:5}
\end{equation}

$t_0(r)$ is another arbitrary function of $r$, representing the BB
singularity surface for which $R(r,t)=0$. One can always choose $t_0(r)=0$ 
at the center $(r=0)$ of the universe by an appropriate translation of the 
$t=$ const. surfaces and describe our universe by the $t>t_0(r)$ part of the 
$(r,t)$ plane, increasing $t$ corresponding to going from the past to the 
future.

Equation (\ref{eq:5}) substituted into equation (\ref{eq:4}) gives:
\begin{equation}
\rho (r,t)={1\over 2\pi G[3t-3t_0(r)-2rt'_0(r)][t-t_0(r)]}  \label{eq:6}
\end{equation}

\subsection{Shell-crossing}

The above expression for $\rho$ leads to two undesirable consequences:\\

1) The energy density goes to infinity not only on the BB surface $t=t_0(r)$, 
but also on the shell-crossing surface:
\begin{equation}
t=t_0(r)+{2\over 3}rt'_0(r)   \label{eq:7}
\end{equation}

2) This energy density presents negative values in the region of the universe 
located between the shell-crossing surface (\ref{eq:7}) and the BB 
singularity, corresponding to $3t - 3t_0(r) - 2rt'_0(r) < 0$ and 
$t - t_0(r) > 0$. One can wonder what does physically mean a negative energy 
density for dust.

Shell-crossing is thus generally considered as a mischief of Tolman-Bondi 
models and physicits usually try to avoid it \cite{HL85}, e.g. by assuming 
$t'_0(r)\leq 0$ for all $r$.

But, as will be developed in next section, we need an increasing BB function 
$t_0(r)$ to solve the horizon problem. Let us hence briefly show how to 
circumvent these two difficulties while keeping  $t'_0(r)>0$.

1) A way out the shell-crossing surface problem is to consider that,
as the energy density increases while reaching its neighbourhood from
higher values of $t$, radiation becomes the dominant component of the universe,
pressure can no more be neglected and the Tolman-Bondi model does no longer
hold.

2) A negative value of $\rho$ proceeds from a negative value of $R'$
in equation (\ref{eq:4}) .

The physical definition of energy density is:
\begin{equation}
\rho \equiv {\delta M\over  \delta V}  \label{eq:8}
\end{equation}

$\delta M$ being the element of mass in an element of volume $\delta V$.

The element of 3-volume corresponding to the flat Tolman-Bondi metric,
i.e. metric (\ref{eq:1}) with $S^2(r,t)=R'^2(r,t)$, is:
\begin{equation}
\delta V=R' R^2\sin \theta dr d\theta d \varphi  \label{eq:9}
\end{equation}
which, when integrated over $\theta$ and $\varphi$, becomes:
\begin{equation}
\delta V=4\pi R' R^2 dr   \label{eq:10}
\end{equation}

As the physical volume $\delta V$ is by convention always positive, equation 
(\ref{eq:10}) possesses a physical meaning only if it is written:
\begin{equation}
\delta V=4\pi |R'| R^2 dr    \label{eq:11}
\end{equation}

And thus, in equation (\ref{eq:4}), one has to replace $R'$ by $|R'|$, which 
gives in equation (\ref{eq:6}):
\begin{equation}
\rho (r,t)={1\over 2\pi G|3t-3t_0(r)-2rt'_0(r)|[t-t_0(r)]}
\label{eq:12}
\end{equation}

However, as, in the following, the light cones of interest never leave
the region situated above the shell-crossing surface in the
$(r,t)$ plane, $3t-3t_0(r)-2rt'_0(r)$ remains positive and equation 
(\ref{eq:6}) holds.

\subsection{Definition of the temperature}

In the course of this letter, we shall be led to use surfaces of constant 
temperature $T$. Since the universe is not homogeneous, there is, at a given 
$t$, no global thermodynamical equilibrium, and $T$ is not readily 
defined. We assume that the characteristic scale of the $\rho$ inhomogeneity 
is much larger than the characteristic length of the photon-baryon 
interaction and that there is always a local thermodynamical equilibrium. 
This enables us to define a local specific entropy $S$ by:

\begin{equation}
S(r) \equiv {k_B n_\gamma (r,t)m_b \over \rho (r,t)}   \label{eq:13}
\end{equation}
where $m_b$ is the baryon mass and $k_B$ the Boltzmann constant.

We then define $T$ by:
\begin{equation}
n_\gamma =a_n T^3   \label{eq:14}
\end{equation}
where  $a_n={2 \zeta(3) k_B^3\over \pi^2 (\hbar c)^3}$

The following expression for $T$ can then be obtained from equations 
(\ref{eq:12}) to (\ref{eq:14}):
\begin{equation}
T(r,t)=\left( {S(r)\over 2\pi G k_B m_b a_n |3t-3t_0(r)-2rt'_0(r)|
[t-t_0(r)]} \right)^{1/3}   \label{eq:15}
\end{equation}

The equation of the $T=$ const. surfaces located after the
shell-crossing surface, i.e. with $3t-3t_0(r)-2rt'_0(r)>0$, is thus the
positive solution of the second order in $t$ equation derived from
equation (\ref{eq:15}):
\begin{equation}
t=t_0(r)+{r\over 3}t'_0(r)+{1\over 3} \sqrt{r^2t'^2_0(r)+
{3S(r)\over 2\pi G k_B a_n m_b T^3}}   \label{eq:16}
\end{equation}

\subsection{The ``centered Earth'' assumption}

In this first approach of a delayed BB solution, the Earth will be
assumed situated sufficiently close to the ``center'' of the universe, so as to
justify the approximation $r_p=0$, the subscript $p$ refering to our
actual location at the present time. We shall comment on this ``center'' of 
the universe in the final discussion.

The value $S_p$ of the entropy function at $(r_p,t_p)$ is $S_p=k_B
\eta_p$, $\eta_p$ being the present local photon to baryon density ratio, 
which is taken to be of order $10^8$.

Remembering that $t_0(r$=$0)$ has been chosen to be zero, one can add to the
specifications of the $t_0(r)$ function:
\begin{eqnarray}
rt'_0 |_{r=0}=0  \nonumber
\end{eqnarray}
to get at $(r=0)$:
\begin{eqnarray}
t_p &\sim & 3.10^{17} s  \qquad \hbox{for} \qquad  T_p=2.73^{\circ} K \nonumber \\
t_{ls} &\sim & 6.10^{12} s \qquad \hbox{for} \qquad  T_{ls}=4 000^{\circ} K \nonumber \
\end{eqnarray}

These values are of the same order of magnitude as in the standard hot
BB model. The nucleosynthesis scenario would thus approximately
be the standard one for the here described universe in the vinicity of
$r=0$, provided the characteristic lenght of the density inhomogeneities
is much larger than the mean free path of the nucleons. This latter condition 
will be discussed in section 5. 

As the light elements abundances predicted by standard cosmology fit rather
well the data observed in our neighbourhood, the choice of the ``center''
of the universe for the location of the Earth seems justified, as far as
the above cited condition obtains.

This good agreement between the observed data and the abundances predicted
by the standard model led Liddle \cite{L95} to adopt the Cosmological 
Principle and thus assert that the FRW metric obtains for the whole universe. 
This was a key-assumption for his tentative proof discussed in our 
Introduction.

But this assumption seems exceedingly narrowing as far as the present available
data have been measured in our direct neighbourhood as compared to
cosmological distances - the today most remote measured abundances are for
deuterium at redshifts $z<5$ \cite{W96}. The ``centered Earth'' assumption 
seems thus enough to complete the game.

\section{Solving the horizon problem}

Light travels from the last scattering surface to a present local observer
on a light cone going from ($r_p=0$, $t_p$) to a 2-sphere ($r_{ls}, t_{ls}$) 
on the last scattering 3-sphere defined by $T=4000^{\circ}$ K.

To solve the horizon problem, it is sufficient to show that this 2-sphere
can be contained inside the future light cone of any ($r=0$, $t>0$) point 
of space-time.

One of the key-points of the reasoning here proposed is a shell-crossing 
surface situated above the BB surface and monotonously increasing 
with increasing $r$, which is always verified if:
\begin{eqnarray}
t'_0(r)&>&0   \qquad   \hbox{for all} \qquad r \nonumber \\
5t'_0(r) + 2rt''_0(r) &>&0   \qquad   \hbox{for all} \qquad r \nonumber
\end{eqnarray}

A $t_0(r)$ function increasing with $r$ implies that the BB ``occured'' at 
later $t$ for larger $r$, hence the evocative ``delayed Big-Bang'' we 
choosed to qualify this singularity.

The model being spherically symmetrical and the null geodesics being radial, 
the relevant light cones are obtained for $\theta =$ const. and 
$\varphi =$ const. in equation (\ref{eq:1}).

Writing $ds^2=0$ in this equation, one gets a differential equation
for the null cones:
\begin{equation}
{dt\over dr}=\pm {R' \over c}   \label{eq:17}
\end{equation}

Substituting above the expression of $R'$ obtained from equation 
(\ref{eq:5}), one finds:
\begin{equation}
{dt\over dr}=\pm {1\over 3c} \left(9G M_0\over 2 \right)^{1/3}
{3t-3t_0(r)-2rt'_0(r) \over [t-t_0(r)]^{1/3}}   \label{eq:18}
\end{equation}

Comparing to equation (\ref{eq:7}), one immediately sees that the curves 
representing the light cones in the $(r,t)$ plane possess an horizontal 
tangent on and only on the shell-crossing surface, where 
$3t-3t_0(r)-2rt'_0(r)$ goes to zero.

The curve $t(r)$ for the past light cone from $(r_p, t_p)$ verifies equation 
(\ref{eq:18}) with the minus sign.

As far as one considers the part of this light cone located after the
shell-crossing surface, and thus after the BB singularity, $3t-3t_0(r)-2r
t'_0(r)$ and $t-t_0(r)$ remain positive and ${dt\over dr}$ is always 
negative. $t(r)$ is a strictly decreasing function of $r$ and the light
cone will have to cross the strictly increasing shell-crossing surface
at a finite point where the derivative of $t(r)$ goes to zero.

On its way to shell-crossing, the null geodesic will cross in turn each
$T=$ const. surface at a finite point.

Let $(r_{ls1}, t_{ls1})$ be the coordinates of the crossing point on the last
scattering surface $T=4000^{\circ}$ K.

Now consider a backward null radial geodesic starting  from any point 
above the shell-crossing surface and directed towards $r = 0$. Its 
equation is a solution of differential equation (\ref{eq:18}) with 
the plus sign.

Its derivative remains positive as long as it does not
reach the shell-crossing surface. If it was to reach this surface before
the ``center'' of the universe, its derivative would go directly
from a positive value to zero, which would imply for the curve of the
light cone an horizontal tangent in the $(r, t)$ plane.

Since we consider models for which the curve representing the shell-crossing 
surface is strictly increasing with $r$, it cannot be horizontally crossed 
from upper values of $r$ and $t$ by a strictly increasing curve. And thus 
one is led to an inconsistency.

This implies that the backward light cone starting from any
point above the shell-crossing surface reachs $r=0$ at $t_c$ without crossing 
this surface, and thus with $t_c>0$.

This statement holds for every light cone issued from any point on
the last scattering surface.

There is thus an infinite number of points $(r=0,$ $t_c>0)$ of which
the future light cone contains the sphere on the last scattering
surface seen today in the CMBR.

Every point on this sphere can be causally connected and the horizon
problem is solved.

\section{Examples of appropriate Big Bang functions}

In previous sections, conditions have been imposed upon the BB 
function $t_0(r)$. They can be summarized as follows:
\begin{eqnarray}
t_0(r=0) &=& 0  \nonumber \\
t'_0(r) &>& 0  \qquad  \hbox{for all} \qquad r  \nonumber \\
5t'_0(r) + 2rt"_0(r) &>& 0 \qquad \hbox{for all} \qquad r \nonumber \\
rt'_0|_{r=0} &=& 0  \nonumber \\
\nonumber\end{eqnarray}

It is easy to verify that the class of functions:
\begin{equation}
t_0(r)=br^n  \qquad b>0  \qquad n>0 \label{eq:19}
\end{equation}
fulfills these conditions.

Another feature imposed upon the model to justify the dust approximation
is that the light cones, for the Tolman-Bondi metric, never leave the matter 
dominated area. This prescription has been tested upon peculiar models of 
the above class with $S(r)=$ const. $= k_B \eta_p$.

A number of light cones were numerically integrated with different values
for $n$ and $b$, in units $c=1$, $t_p={9GM_0\over 2}$.

It has been in particular found that:

(1) For  $n=1$, the backward null geodesics starting from 
$(r_{ls1}, t_{\rm ls1})$ reach $r=0$ without leaving the matter dominated 
area (approximately delimited near $r=0$ by 
$T=T_{\rm eq}= 10^5$ $ ^{\circ} K$) provided $b$ is kept larger than about 
$10^{12}s$.

(2) for $n=2$, an analogous condition holds, the limiting value for $b$ 
being about $10^{14}s$.

Figures 1 and 2 show the case $n=2$, $b=5$x$10^{14}$s for which 
$t_c=2.18~10^{12}$s.

\section{Discussion and conclusion}

Using a delayed BB universe, the horizon problem has been solved
for a class of simple models fulfilling some restricting conditions.
Further work will be necessary to discriminate between these conditions
which are generic and which are only generated by the assumptions
made for simplification purpose.

For instance, the dust approximation used in this letter has been
retained to allow analytical calculations. It was shown in section 4 that 
this dust choice provided constraints upon the BB function $t_0(r)$.

The behaviour of light cones in the radiation dominated region is thus
an appealing issue for future work. If it could be proven that the geometry 
of this region does not bend the light cones such as to have them reach the
BB surface before $r=0$, then the above constraints could be discarded.

One could therefore consider the a priori interesting case of a BB function, 
fulfilling conditions summarized at the beginning of section 4, but arbitrarily
close to the FRW $t_0(r)=0$. 

In this case, the ``unnatural'' prescription ``centered Earth'' 
would no more be needed as the conditions for standard nucleosynthesis 
could be verified for an observer located at arbitrary values of $r$.

This would imply that the characteristic lenght of the density 
inhomogeneities, written for the radiation dominated model, at the Earth 
location, is much smaller than the mean free path of the nucleons.  
A constraint would thus appear on $t_0(r)$, i.e. limiting $b$ to small values 
compatible with the ``close to FRW'' assumption.

The authors are well aware of a potential difficulty of the present model, 
namely to put the observer near the ``center'' of the universe. In addition to 
the fact that such a location is not forbidden by scientific but only by 
philosophical principles (which they do not accept), they want to stress that 
the present model is only a first ``toy-model''. They hope to build, in the 
future, less simple models, getting rid of this prescription.

In a recent work, they have shown that a delayed BB of type 
$t_0(r) = br$ can reproduce the observed dipole and quadrupole in the 
CMBR anisotropies - a first version of this work has been submitted for 
publication \cite{SC98} with a decreasing BB function (negative values for 
$b$), but it is easy to see that the same results hold for $b>0$.

It comes out from this work that to any given value of the location $r_p$ 
of the observer corresponds a value of $b$ for which the observed data are 
reproduced. $b$ is all the smaller as $r_p$ is larger.

A reliable model of universe of this kind could thus get rid of the 
``centered Earth'' assumption, provided $b$ should be 
sufficiently small. This implies that the null geodesics, if causally 
connected, should be so in the radiation dominated region.

Now, why should we feel unconfortable with the idea that we could be 
located near the ``center'' of the universe? Following Ellis, Maartens and 
Nel \cite{E78,EMN78}, who also dared assume such an ``unnatural'' 
prescription in their Static Spherically Symmetric (SSS) model of universe, 
one can claim that this is no more (un)reasonable than the belief in a 
Cosmological Principle. The purpose is not to put the observer at the 
``center'' a priori, but to answer the question as it was put forward by 
these authors in their cited papers: ``Given a universe model of the type 
proposed, where would one be likely to find life like that we know on 
Earth?''. The answer of Ellis is: ``where conditions are favorable for life 
of this kind ... near the center, where the universe is cool.''
   
The flat universe approximation, even if it seems more physically justified, 
can also be discussed. If our universe would be proved not
so flat as it seems to be, the study of the open (closed ?) case would
become necessary.

However, even if the here proposed class of models appears as a
restrictive answer to the horizon problem, it might, with some easy
to conceive improvements, equally account for structure formation and all
scales anisotropies of the CMBR, which could otherwise proceed from a 
topological defects like mechanism. This will be the purpose of other 
work to come.

Inflation was, from the beginning of its success story, equally aimed at
solving the flatness and monopole problems.

As it is a mere product of Friedmann's equations, the flatness problem only 
pertains to FRW universes and is thus irrelevant for the class of models 
here proposed.

As for the monopole problem, delayed BB without inflation only implies 
that topological defects theories leading to a production of local 
stable monopoles are ruled out.

This letter is a first attempt to show that the delayed
BB scenario, as it is a natural and simple way to solve the
problems of standard cosmology while keeping its best successfull predictions,
is worth spending time and efforts to bring it to at least as worthy a
paradigm as any other on the market place.

The last point to emphasize is the following. In the years to come, two
satellite boarded missions, MAP to be launched by NASA and Planck to be
launched by ESA, will be dedicated to a high-resolution maping of the CMBR
anisotropies. One of their main purposes is to provide a test of cosmological 
theories and an estimation of cosmological parameters.

Number of recent papers attempt to show how the values of these
parameters could be determined by an analysis of the data thus obtained,
see e.g. \cite{B94}, \cite{J96} and \cite{BET97}. In these papers, the BB 
function $t_0(r)$ is always implicity or explicity set to a constant value
over the spatial coordinate $r$, and the cosmological parameters considered
are those pertaining to a universe with FRW background.

To be complete, the analysis of these future data will also have to be 
performed in the light of the present results.

\end{document}